\documentclass[ba]{imsart}

\RequirePackage{amsthm,amsmath,amsfonts,amssymb}
\RequirePackage[authoryear]{natbib}%% uncomment this for author-year citations
\RequirePackage[colorlinks,citecolor=blue,urlcolor=blue,backref=page,backref=page]{hyperref}
\RequirePackage{graphicx}
\usepackage{subcaption}
\usepackage{booktabs}

\pubyear{2025}
\arxiv{2509.22901}
\volume{TBA}
\issue{TBA}
\firstpage{1}
\lastpage{1}

\startlocaldefs
\theoremstyle{plain}

\theoremstyle{definition}

\theoremstyle{remark}

\endlocaldefs

\begin{document}

\begin{frontmatter}
\title{An exploration of sequential Bayesian variable selection - A comment on Garc\'{i}a-Donato et al. (2025). ``Model uncertainty and missing data: An objective Bayesian perspective''}
%\title{A sample article title with some additional note\thanksref{t1}}
%\runtitle{}
%\thankstext{T1}{A sample additional note to the title.}

\begin{aug}
%%%%%%%%%%%%%%%%%%%%%%%%%%%%%%%%%%%%%%%%%%%%%%%
%% Only one address is permitted per author. %%
%% Only division, organization and e-mail is %%
%% included in the address.                  %%
%% Additional information can be included in %%
\author[A]{\fnms{Sebastian}~\snm{Arnold}\ead[label=e2]{Sebastian.Arnold@cwi.nl}\orcid{0000-0003-4021-5295}}
\and
\author[A]{\fnms{Alexander}~\snm{Ly}\ead[label=e1]{Alexander.Ly@cwi.nl}\orcid{0000-0003-3925-3833}},
%%%%%%%%%%%%%%%%%%%%%%%%%%%%%%%%%%%%%%%%%%%%%%
%% Addresses                                %%
%%%%%%%%%%%%%%%%%%%%%%%%%%%%%%%%%%%%%%%%%%%%%%
\address[A]{Machine Learning,
Centrum Wiskunde \& Informatica\printead[presep={,\ }]{e1}}
%
%\address[B]{Mathematical Institute,
%Leiden University\printead[presep={,\ }]{e3}}
\runauthor{S. Arnold and A. Ly}
\end{aug}

%\begin{abstract}
%The abstract should summarize the contents of the paper. It
%should be clear, descriptive, self-explanatory and not longer
%than 200 words. It should also be suitable for publication in
%abstracting services. Please avoid using math formulas as
%much as possible.
%
%This is a sample input file.  Comparing it with the output it
%generates can show you how to produce a simple document
%of your own.
%\end{abstract}
%
%\begin{keyword}[class=MSC]
%\kwd[Primary ]{00X00}
%\kwd{00X00}
%\kwd[; secondary ]{00X00}
%\end{keyword}
%
%\begin{keyword}
%\kwd{First keyword}
%\kwd{second keyword}
%\end{keyword}

\end{frontmatter}

Congratulations to Garc\'{i}a-Donato, Castellanos, Cabras, Quir\'{o}s, and Forte (\citeyear{garcia2025model};  henceforth GCCQF) for their valuable work extending the Bayesian variable selection methodology to handle missing data, which significantly broadens its applicability to real-world scenarios. The paper offers a valuable basis for further discussion and we are pleased by the invitation to write this comment.

Our comment explores a further extension of the proposed methodology. Specifically, we consider the sequential setting where (potentially missing) data accumulate over time, with the goal of continuously monitoring statistical evidence, as opposed to assessing it only once data collection terminates. To this end, we replicated the first setting of their Experiment 1 using data generated under the linear regression model
\begin{align}
y_{i} = x_{i1} + 2 x_{i2} + x_{i6} + 2 x_{i7} + \epsilon_{i}, \quad \epsilon_{i} \sim \mathcal{N}(0, \sigma^{2}), \text{ where } \sigma^{2} = 2.5 ,
\end{align}
where the four active variables \( x_{i1}, x_{i2}, x_{i6}, x_{i7} \) are correlated with 6 other covariates, all jointly normally distributed. In each replication, 100 outcomes \( y \in \mathbb{R} \) and potential covariates \( \boldsymbol{x} \in \mathbb{R}^{p}\), where \( p=10 \), were generated. As in the main text, 40\% of the covariates were made missing at random. Instead of running their \texttt{R} procedures only at \( n=100 \), we did so repeatedly at \( n=19, 20, \ldots, 100 \), resulting in \( t=1, \ldots, 82 \) imputed covariate sets and posterior inclusion probabilities (the used imputation method needs a minimum number of data points to run properly, and we could, thus, not track the posterior probabilities for \( n<19 \)). As opposed to the original setting, we did decrease the number of mice imputations from 500 to 50, but this did not qualitatively change the reported results at \( n=100 \). 

\section{Exploring sequential Bayesian variable selection}
\begin{figure}[htbp]
    \centering
    \begin{subfigure}[b]{0.48\textwidth}
        \includegraphics[width=\textwidth]{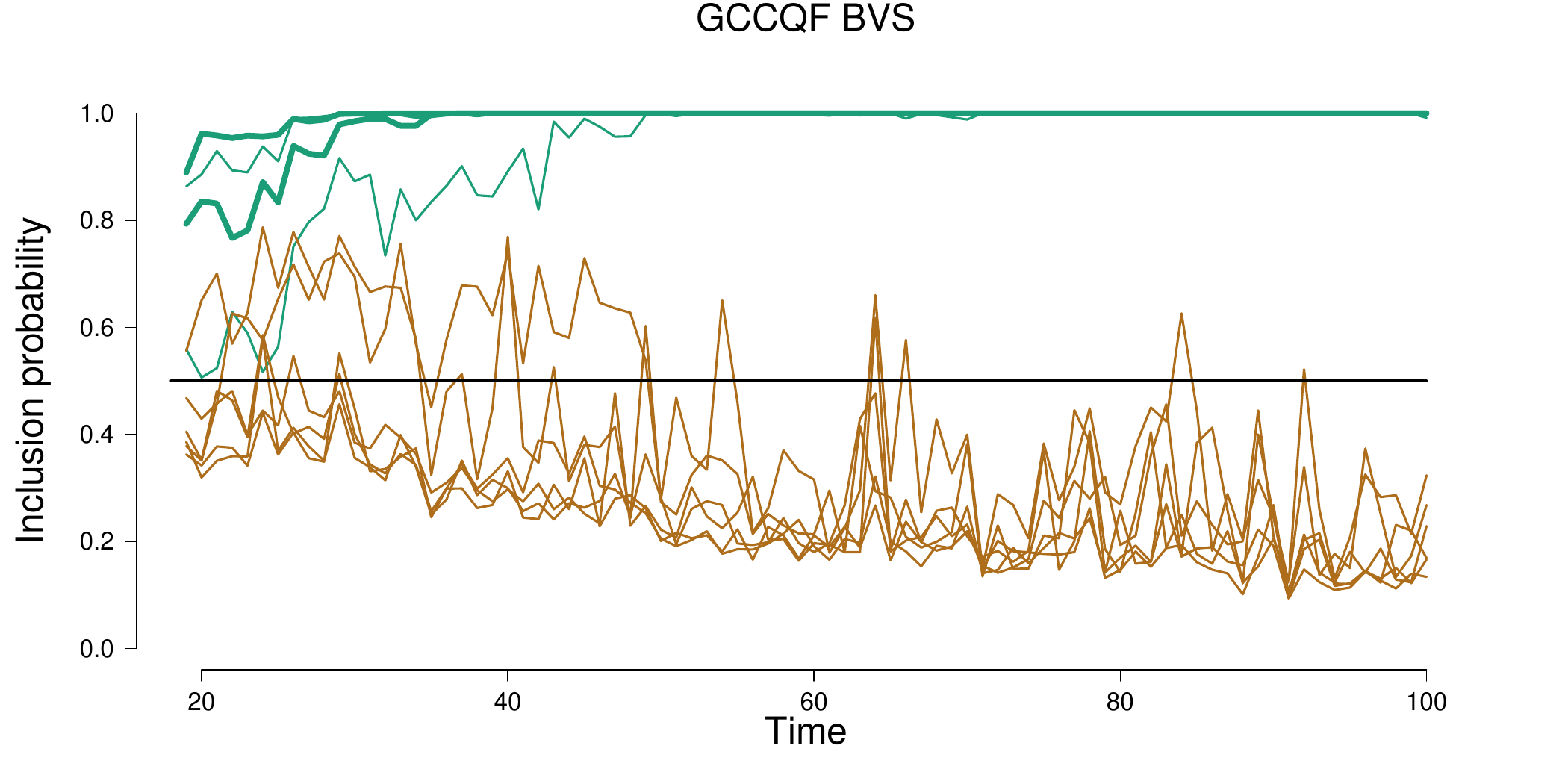}
        \label{fig19Bvs}
    \end{subfigure}
    \hfill
    \begin{subfigure}[b]{0.48\textwidth}
        \includegraphics[width=\textwidth]{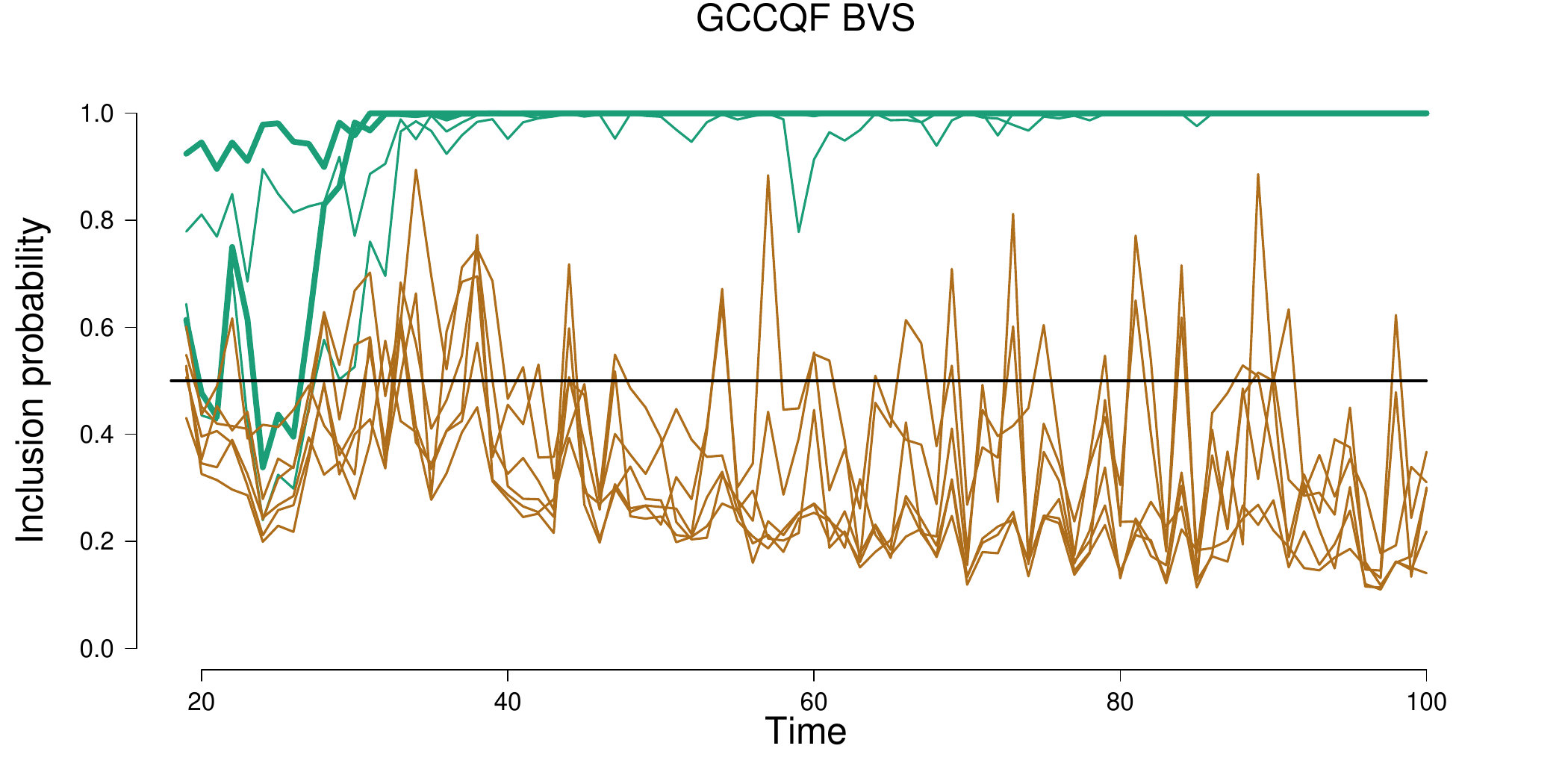}
        \label{fig76Bvs}
    \end{subfigure}

    \vspace{0.5cm} % Add vertical spacing between rows

    \begin{subfigure}[b]{0.48\textwidth}
        \includegraphics[width=\textwidth]{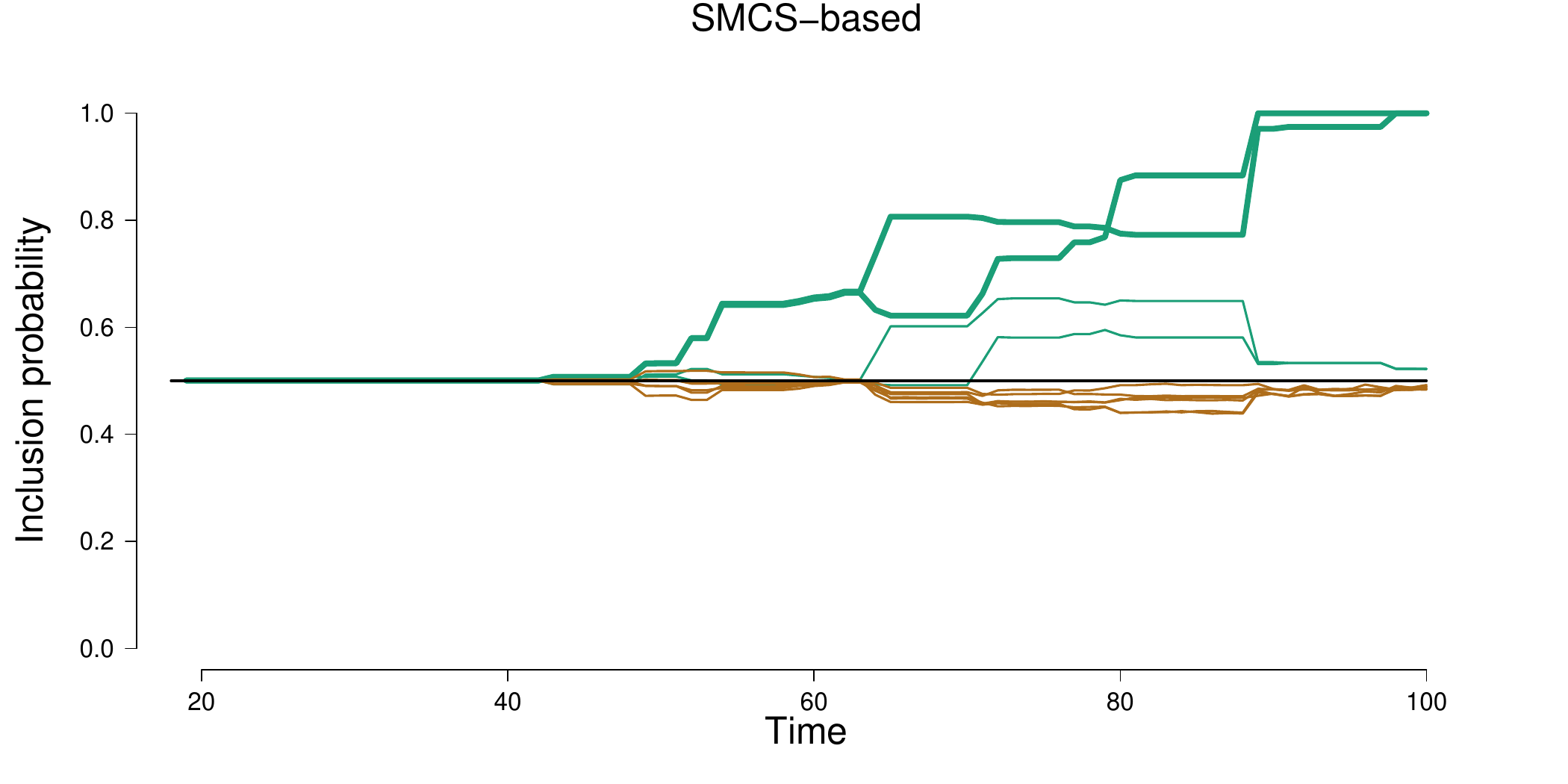}
        \label{fig19Mixed}
    \end{subfigure}
    \hfill
    \begin{subfigure}[b]{0.48\textwidth}
        \includegraphics[width=\textwidth]{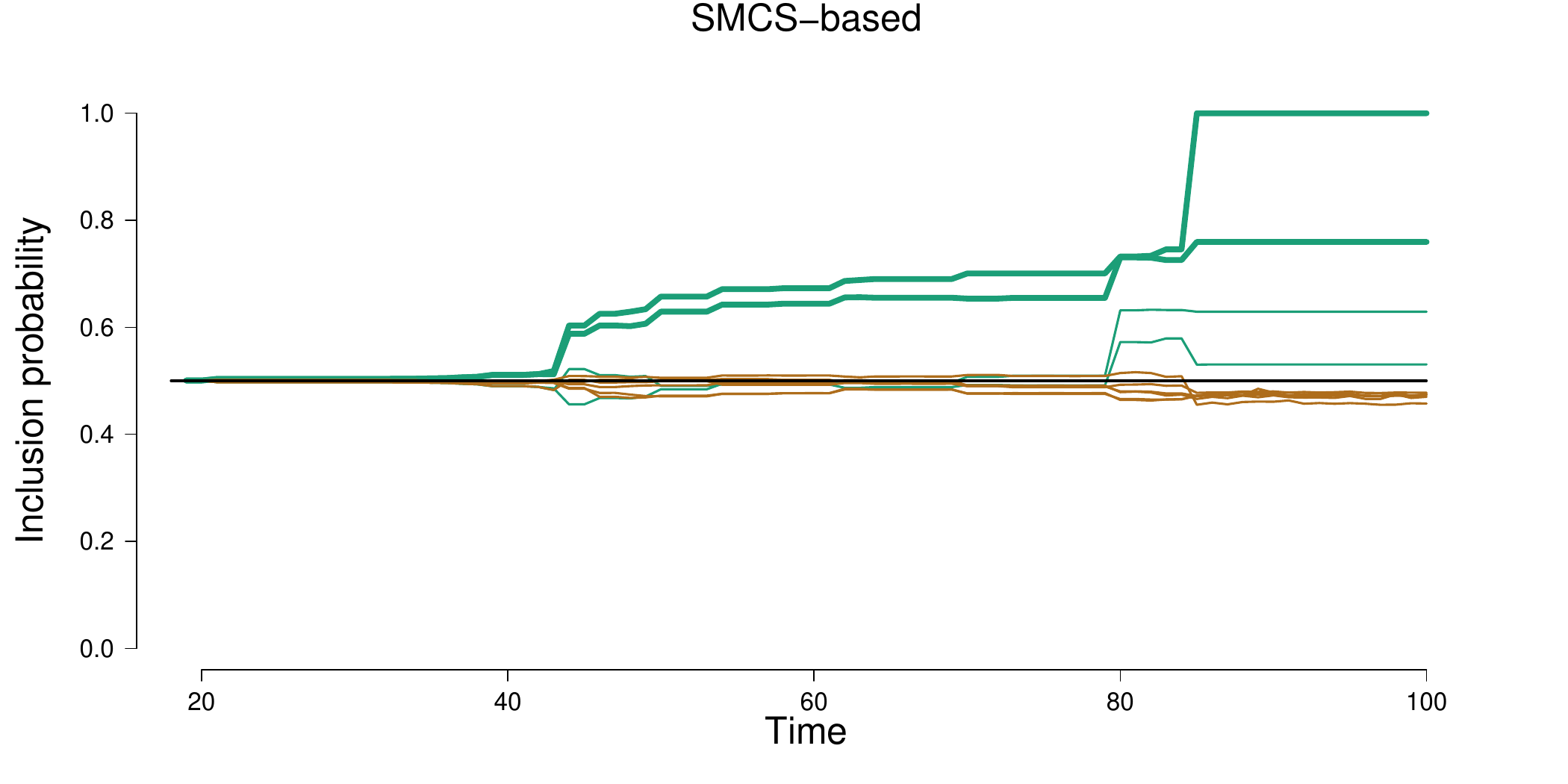}
        \label{fig76Mixed}
    \end{subfigure}

    \vspace{0.5cm} % Add vertical spacing between rows

    \begin{subfigure}[b]{0.48\textwidth}
        \includegraphics[width=\textwidth]{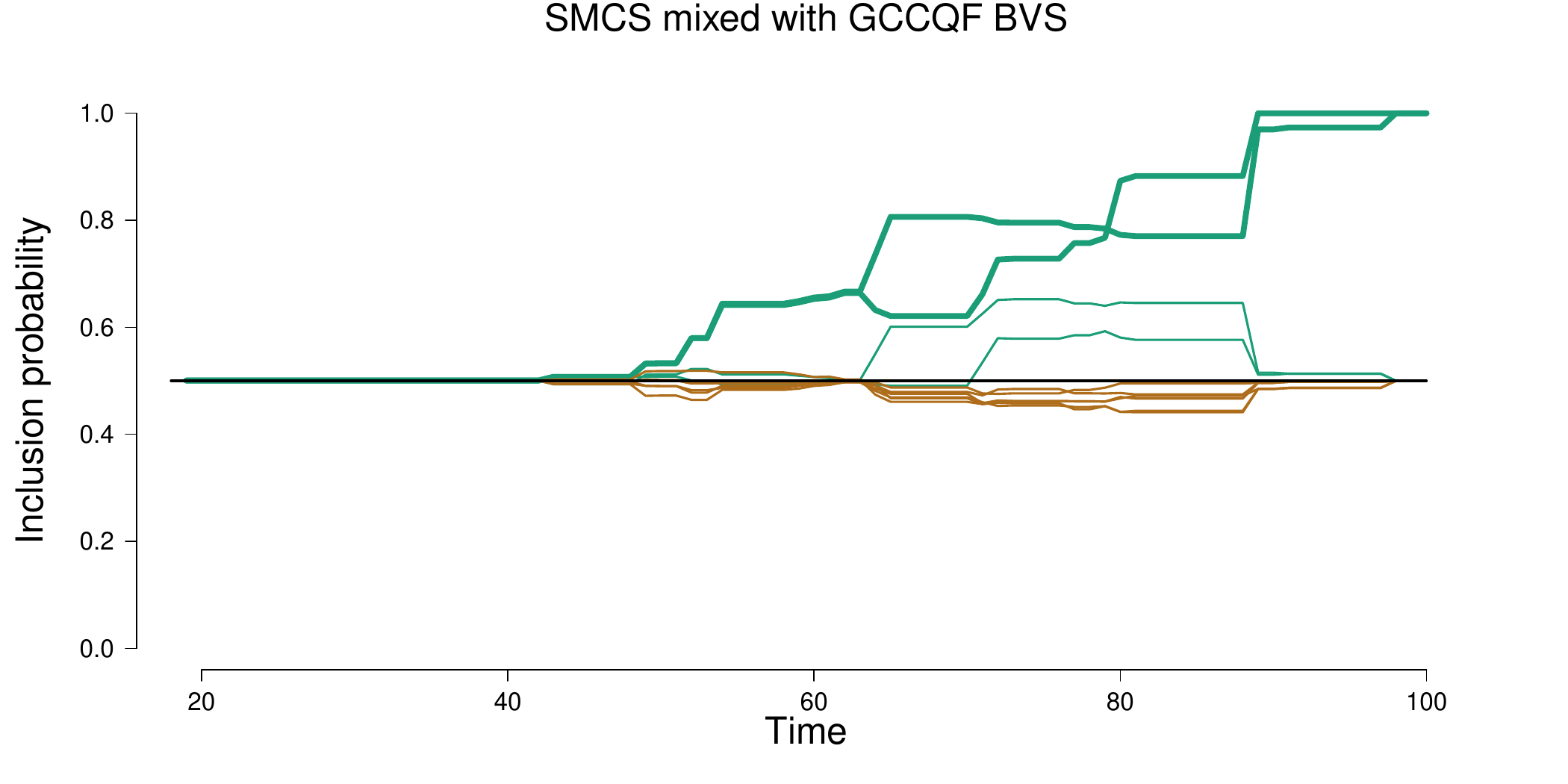}
        \label{fig19Smcs}
    \end{subfigure}
    \hfill
    \begin{subfigure}[b]{0.48\textwidth}
        \includegraphics[width=\textwidth]{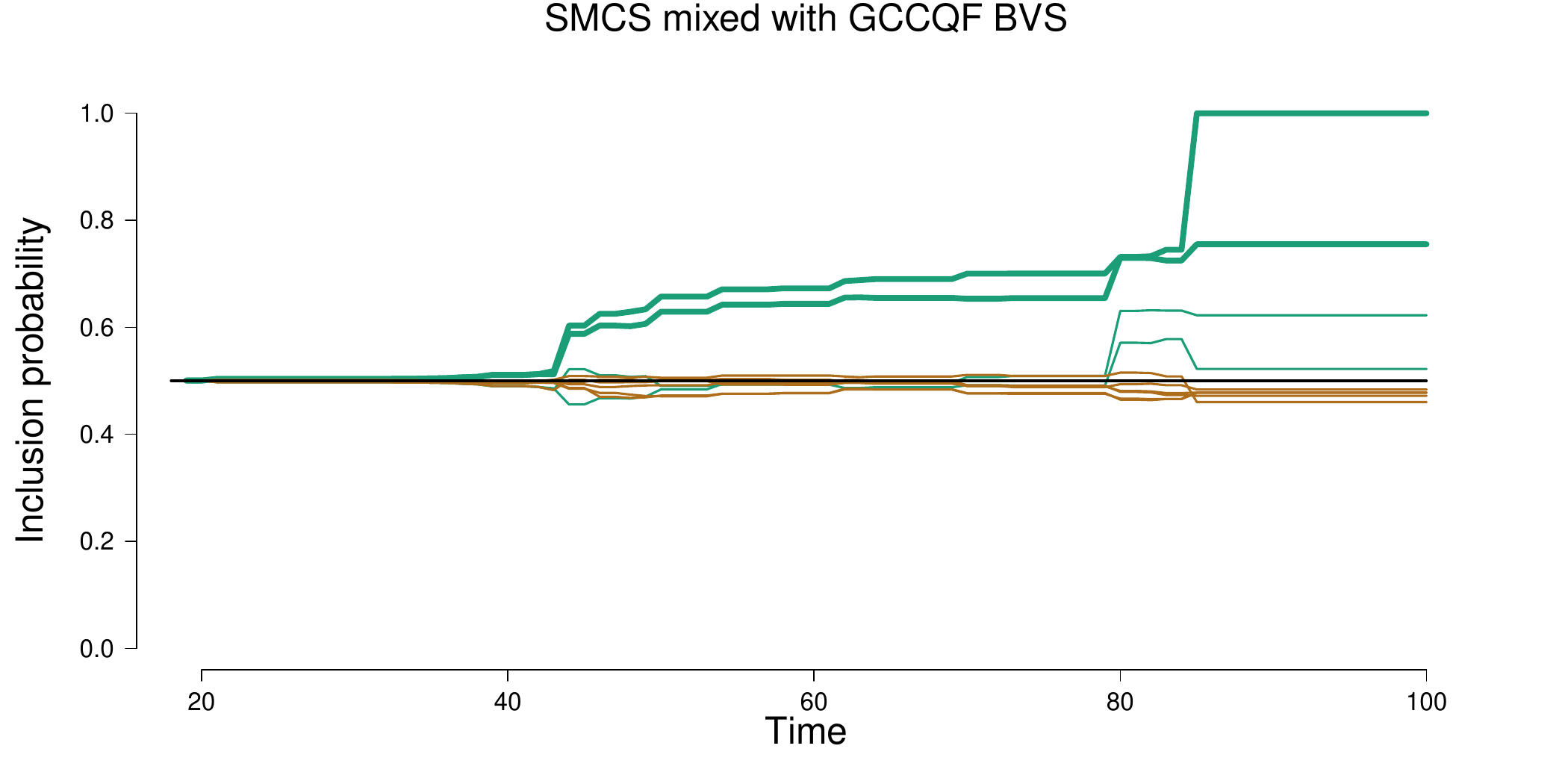}
        \label{fig76Smcs}
    \end{subfigure}
    \caption{Posterior inclusion probabilities of the active (green) and inactive (brown) covariates over time for two runs of the simulation with respect to the sequential application of GCCQF (top), the SMCS-based approach (bottom), and the mixture approach (middle), where \( \alpha = 0.1 \) and \( \lambda = 1/(8 \varsigma^{2}) \approx 0.3 \), for \( \varsigma=0.65\leq \sigma \). Thick lines indicate the active covariates \( x_{2} \) and \( x_{7} \) with larger regression coefficients.}
    \label{figSequential}
\end{figure}

In the simulation study we applied the sequential procedure to \( 100 \) data sets, and the top row of Figure~{\ref{figSequential}} shows the result of the GCCQF procedure for simulation 19 and 76, respectively. %
The posterior inclusion probabilities of the active and inactive covariates are depicted as green and brown curves, respectively. The top left panel shows that the inclusion probabilities of the active covariates remained above \( 0.5 \) throughout, but that the inactive covariates are \( 17 \) times misidentified as being active during data collection. The top right panel shows \( 4 \) and \( 46 \) misidentifications of the active and inactive covariates, respectively. To stabilise inference, we combined an initial approach based on sequential model confidence sets (bottom row of Figure~\ref{figSequential}) with the GCCQF procedure, yielding the results shown in the second row of Figure~\ref{figSequential}.

\section{Sequential model confidence sets}
In our study, we use sequential model confidence sets, as proposed by \cite{arnold2024sequential}, as a meta-algorithm for sequential model selection. This method compares the \( m=2^{p} \) candidate models represented by binary vectors \( \boldsymbol{\gamma}^{i} = (\gamma_{1}^{i}, \dots, \gamma_{p}^{i}) \in \{ 0,1 \}^{p} \), \( i \in [m]:=\{1, \dots, m \} \), based on their predictive performance. At time \( t \in \mathbb{N} \), we observe \( y_{t} \) and compute the loss of model \( i \), denoted by \( L_{i, t} = \ell( \hat{y}_{i,t}, y_{t}) \), where \( \hat{y}_{i,t} \) is the model's prediction and \( \ell \) some negatively oriented loss function. The loss differences \( d_{ij, t} = L_{i, t} - L_{j, t} \leq 0 \), iff model \( i \) outperforms model \( j \) at time \( t \). Proposition~{3.3} in \cite{arnold2024sequential} implies that, if (i) the data are generated by some model \( i^{\star} \in [m] \) under consideration, and (ii) the loss differences are conditionally sub-exponential with tuning parameter \( \lambda \geq 0 \), then the collection \( \widehat{\mathcal{M}}_{t} \) consisting of all models \( i \in [m] \) for which the following holds %
\begin{align}
\label{eqEProcesses}
E_{i, t} = \sup_{r \leq t} \frac{1}{m-1} \sum_{j \neq i} \exp \left\{ \lambda \left(\sum_{s=1}^{r} d_{ij, s}\right) - r/8\right\}  \leq 1/\alpha, \quad t \in \mathbb{N},
\end{align}
forms a so-called \emph{sequence of model confidence sets} (SMCS) or simply \emph{sequential model confidence sets} (SMCSs) at level \( \alpha \in (0,1) \), which means that
\begin{align}
\label{eqDefCoverage}
\mathbb{P}(\forall t \geq 1: i^\star \in \widehat{\mathcal{M}}_{t}  ) \geq 1-\alpha .
\end{align}
That is, the SMCS guarantees simultaneous coverage of the true data-generating model \( i^{\star}\) for all times \( t \) with high probability. Property \eqref{eqDefCoverage} is equivalent to the statement that the probability of the true model ever being excluded from the SMCS is at most \( \alpha \):
\begin{align}
\mathbb{P}(\exists t\geq 1:  i^{\star} \notin \widehat{\mathcal{M}}_{t}  ) \leq \alpha .
\end{align}
This bound is known as Ville's inequality and generally holds for exceedance probabilities of \( E \)-processes, which are the key objects in safe anytime-valid inference, see for instance \citet{grunwald2024safe}, \cite{howard2021time}, and in simplified form \cite{ly2025tutorial}. Inspired by the Bayesian variable selection approach advocated by the authors, we can derive the SMCS-based inclusion probability of covariate \( k=1, \dots, p \) as
\begin{align}
\label{eqInclusionProbabilitySmcs}
    \hat{p}_{t}^{\text{SMCS}}(\gamma_{k} = 1 \mid \boldsymbol{x}_{1}, y_{1}, \ldots, \boldsymbol{x}_{t}, y_{t}) = \frac{\lvert \{i \in  \widehat{\mathcal{M}}_t \mid \gamma_{k}^i = 1\}\rvert}{\lvert \{i \in  \widehat{\mathcal{M}}_t \} \rvert}, \quad t\in \mathbb{N} .
\end{align}
Raw counting results from \( \widehat{\mathcal{M}}_{t} \) being a frequentist construct that either does, or does not, contain the true data generating model. Under repeated use, this procedure is expected to -- at all times -- cover the truth in \( 1- \alpha \)\% of the cases, unlike Bayesian posterior model probabilities, which represent gradually updated beliefs. 

\section{Exploring SMCS inclusion probabilities}
Our initial attempt to use the predictions of each model assessed by the \( \ell_{2} \) loss fits the general theory, but did unfortunately not yield the results we were searching for, perhaps caused by sub-optimally chosen parameters \( \lambda \) and \( \alpha \). Eventually, we used the logarithm of the (imputed) GCCQF Bayes factors to score each model. In particular, the ``loss'' at time \( t \) of model \( i \in [m] \) was computed as 
\begin{align}
L_{i, t} := \frac{1}{m-1} \sum_{j \neq i} \log \text{BF}_{ji, t}, \text{ where } \text{BF}_{ji, t} := \frac{m_{\boldsymbol{\gamma}^{j}}(y_{1}, \ldots, y_{t} \mid \boldsymbol{x}_{1}, \ldots, \boldsymbol{x}_{t})}{m_{\boldsymbol{\gamma}^{i}}(y_{1}, \ldots, y_{t} \mid \boldsymbol{x}_{1}, \ldots, \boldsymbol{x}_{t})}. 
\end{align}
The inclusion probabilities for the two simulation runs are depicted in the bottom row of Figure~{\ref{figSequential}}. Compared to the sequential application of the GCCQF procedure, there is a significant reduction in both the fluctuations and number of crossings through the critical threshold \( 0.5 \) (from 17 to 5 in the left, and from 50 to 9 in the right column). For the simulation run on the right, this increase in stability came at no cost of accuracy at \( n=100 \), thus, \( t=82 \). However, the SMCS based inclusion probabilities do misclassify \( x_{1} \) and \( x_{6} \) as inactive from about \( n=98 \), thus, \( t=80 \), onward in the left column. At this time point about 256 models remain in \( \widehat{\mathcal{M}}_{t} \), and the results show that these weak active covariates appear in exactly half of them. 

To borrow strength from the Bayesian variable selection approach, we explored two methods. The first simply sets the GCCQF posterior model probabilities for model \( i \in [m] \) at time \( t \) to zero, whenever \( i \not \in \widehat{\mathcal{M}}_{t} \). After renormalisation, we then qualitatively recovered the standard Bayesian variable selection approach, which suggests close alignment of the posterior model probabilities and sequential model confidence sets. The second method mixes the GCCQF and SMCS inclusions probabilities, where the last one is proportionally weighted by \( | \widehat{\mathcal{M}}_{t} | / m \), the relative number of surviving models at time \( t \). This implies that the posterior model probabilities will dominate once a significant number of candidate models are removed from \( \widehat{\mathcal{M}}_{t} \). The latter typically occurs for large \( t \), when the Bayesian inclusions probabilities are hopefully stabilised. The middle row of Figure~{\ref{figSequential}} shows that this mixed approach identifies the active coefficients more accurately, while at the same time leading to more stable inference. %
\begin{table}[htbp]
    \centering
    \caption{Average number of crossing of each covariate}
    \label{tabAverageCrossings}
    \begin{tabular}{lcccccccccc}
        \toprule
        \textbf{Method} & \( \beta_{1} \) & \( \beta_{2} \) & \( \beta_{3} \) & \( \beta_{4} \) & \( \beta_{5} \) & \( \beta_{6} \) & \( \beta_{7} \) & \( \beta_{8}  \)& \( \beta_{9} \) & \( \beta_{10} \) \\
        \midrule
        GCCQF BVS & 1.08 & 0.09 & 3.82 & 2.65 & 2.96 & 2.46 & 0.16 & 3.79 & 4.35 & 3.96 \\
        Mixed & 0.47 & 0.05 & 0.88 & 0.74 & 0.63 & 0.82 & 0.15 & 0.98 & 1.29 & 0.93 \\
        SMCS based & 0.46 & 0.05 & 0.8 & 0.66 & 0.7 & 0.75 & 0.15 & 0.87 & 1.02 & 0.94 \\
        \bottomrule
    \end{tabular}
\end{table}

\begin{figure}
\centering
\includegraphics[width= \linewidth]{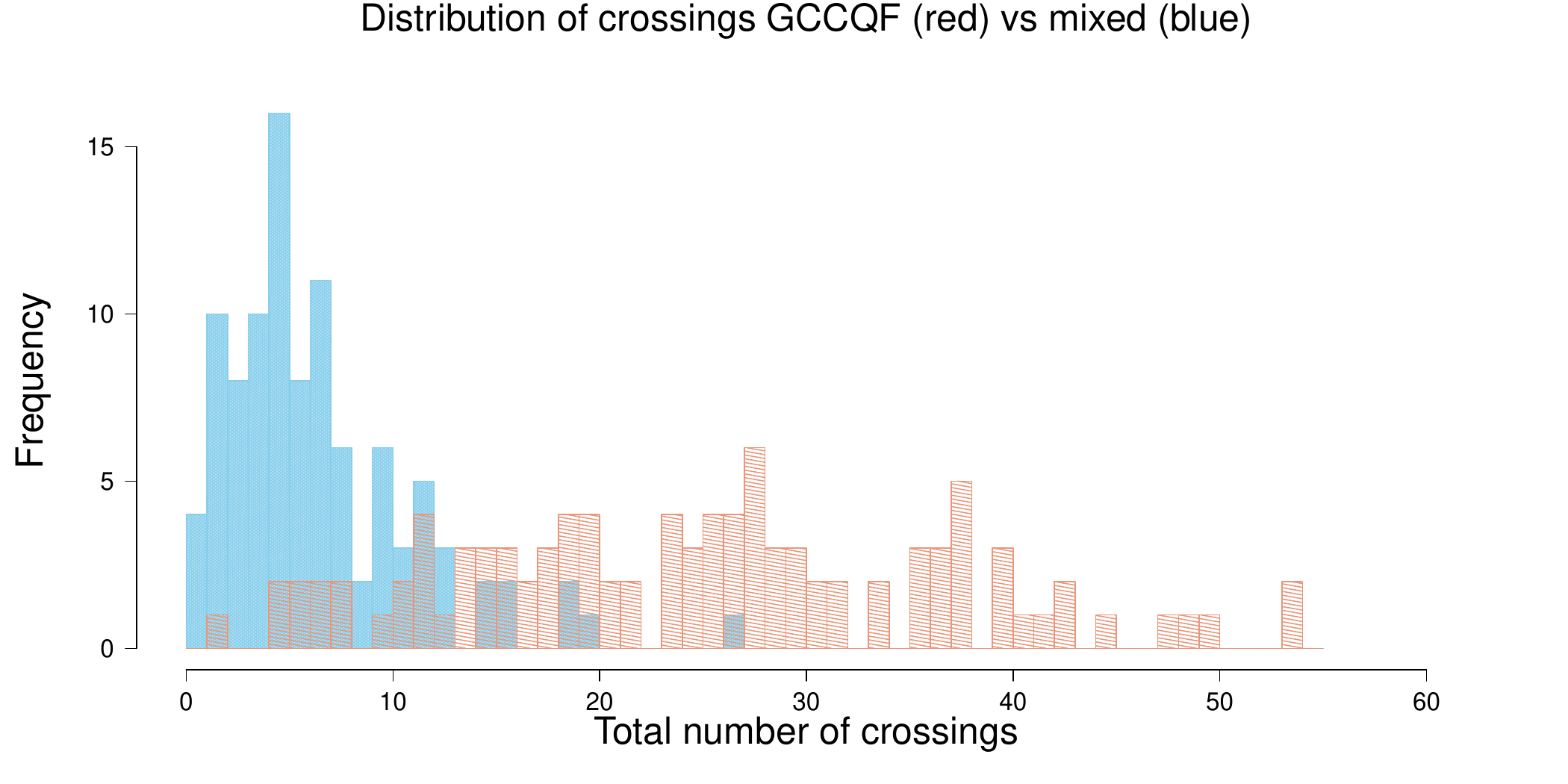}
\caption{The averaged total number of crossings through the critical value \( 0.5 \) for the posterior inclusion probabilities (red) and the mixture approach (blue).} %
\label{figTotalCrossings}
\end{figure}

\begin{table}[htbp]
    \centering
    \caption{The relative frequency of each covariate being included at time \( n=100 \) (\( t=82 \)).}
    \label{tabRelativeFrequency}
    \begin{tabular}{lcccccccccc}
        \toprule
        \textbf{Method} & \( \beta_{1} \) & \( \beta_{2} \) & \( \beta_{3} \) & \( \beta_{4} \) & \( \beta_{5} \) & \( \beta_{6} \) & \( \beta_{7} \) & \( \beta_{8}  \)& \( \beta_{9} \) & \( \beta_{10} \) \\
        \midrule
        GCCQF BVS & 1 & 1 & 0.03 & 0.01 & 0.04 & 0.99 & 1 & 0.02 & 0.09 & 0.05 \\
        Mixed & 0.98 & 1 & 0.04 & 0.10 & 0.02 & 0.97 & 1 & 0.25 & 0.19 & 0.10 \\
        SMCS based & 0.86 & 1 & 0.08 & 0.13 & 0.02 & 0.84 & 1 & 0.26 & 0.20 & 0.11 \\
        \bottomrule
    \end{tabular}
\end{table}
The stability improvements hold consistently across all 100 replicated data sets (Table~\ref{tabAverageCrossings}). While this gain in stability incurs minimal cost in identifying active covariates (Table~\ref{tabRelativeFrequency}), it does increase the misclassification rate of inactive covariates as active. Figure~{\ref{figTotalCrossings}} visualises the decrease in variability of the total number of crossings through the critical value \( 0.5 \). %
\section{Concluding comments and further discussion}
With this note we do not want to critique the contribution of GCCQF, but rather highlight the many interesting challenges involved with sequential decision making. The approach we took here was fully exploratory, as we honestly do not know what can be reasonably expected from a sequential variable selection method. It is unrealistic to demand such method to perfectly identify the (in)active covariates as (in)active as soon as possible, and retain those classifications throughout. But what is realistic? The increase in stability based on mixing the SMCS and GCCQF inclusion probabilities was significant, but we do not know whether it is an optimal procedure. In fact, we do not even know if our choice of \( \lambda \) and \( \alpha \) is optimal for our mixed procedure. The chosen \( \lambda \approx 0.3 \) works well for the underlying true \( \beta \), and we can make it adaptive to other \( \beta \)s. For instance, with a prior distribution, or using a prequential plugin approach. The role of \( \alpha \) requires more explorations and thought. It is also unclear whether the cost of increased misclassification of inactive covariates, as shown in Table~{\ref{tabRelativeFrequency}}, is necessary. Furthermore, it remains unclear how the theoretical guarantees of SMCS, derived in \citet{arnold2024sequential}, translate from the level of models to the derived inclusion probabilities \eqref{eqInclusionProbabilitySmcs}. The use of SMCSs, however, is not totally arbitrary as admissible anytime-valid inference should rely on \( E \)-processes \citep{ramdas2020admissible}, and SMCSs provide a natural candidate for sequential model selection. A central result in the theory of safe sequential inference also suggests that (log) optimal procedures should be Bayesian in nature \citep{grunwald2024safe,larsson2025numeraire}. Our meta-approach moves out of this realm, and we wonder whether the authors can see avenues in making the approach Bayesian again. Perhaps they suggest a different approach to sequential Bayesian variable selection over our naive approach. And would an increase in the rate of misclassification of an inactive covariate be worth the cost of more stable inference?

%%%%%%%%%%%%%%%%%%%%%%%%%%%%%%%%%%%%%%%%%%%%%%%
%%% Acknowledgements                         %%
%%% should be provided in the                %%
%%% Acknowledgements section.                %%
%%%%%%%%%%%%%%%%%%%%%%%%%%%%%%%%%%%%%%%%%%%%%%%
\begin{acks}[Acknowledgments]
The authors would like to thank Stefano Cabras and Gonzalo Garc\'{i}a-Donato for their code, upon which this work builds. They also acknowledge Udo Boehm for valuable discussions and inputs. 
\end{acks}

%%%%%%%%%%%%%%%%%%%%%%%%%%%%%%%%%%%%%%%%%%%%%%
%% Funding information, if any,             %%
%% should be provided in the                %%
%% funding section.                         %%
%%%%%%%%%%%%%%%%%%%%%%%%%%%%%%%%%%%%%%%%%%%%%%
\begin{funding}
This research was funded by the Dutch Research Council (NWO) through the VENI fellowship grant "Increasing Scientific Efficiency with Sequential Methods" (VI.Veni.211G.040) awarded to AL. SA acknowledges funding by the ERC advanced grant (101142168) awarded to Peter Gr\"{u}nwald. 
\end{funding}

%%%%%%%%%%%%%%%%%%%%%%%%%%%%%%%%%%%%%%%%%%%%%%
%% Supplementary Material, including data   %%
%% sets and code, should be provided in     %%
%% {supplement} environment with title      %%
%% and short description. It cannot be      %%
%% available exclusively as external link.  %%
%% All Supplementary Material must be       %%
%% available to the reader on Project       %%
%% Euclid with the published article.       %%
%%%%%%%%%%%%%%%%%%%%%%%%%%%%%%%%%%%%%%%%%%%%%%%
%\begin{supplement}
%\stitle{Title of Supplement A}
%\sdescription{Short description of Supplement A.}
%\end{supplement}
%\begin{supplement}
%\stitle{Title of Supplement B}
%\sdescription{Short description of Supplement B.}
%\end{supplement}

\bibliographystyle{ba}
\bibliography{biblio}

\end{document}